\documentclass[fleqn,10pt]{wlscirep}
\title{TomoReal: Tomographic Displays}

\usepackage{multirow}
\usepackage{colortbl}
\usepackage{tabulary}
\usepackage{tabularx}
\usepackage{colortbl}
\usepackage{amsmath}

\definecolor{lightgray}{gray}{0.9}

\newcolumntype{L}[1]{>{\raggedright\arraybackslash}p{#1}}
\newcolumntype{C}[1]{>{\centering\arraybackslash}p{#1}}
\newcolumntype{R}[1]{>{\raggedleft\arraybackslash}p{#1}}
\usepackage{booktabs} 

\author[]{Seungjae Lee}
\author[]{YoungJin Jo}
\author[]{Dongheon Yoo}
\author[]{Jaebum Cho}
\author[]{Dukho Lee}
\author[*]{Byoungho Lee}

\affil{School of Electrical and Computer Engineering, Seoul National University, Gwanak-Gu Gwanakro 1, Seoul, 08826, Republic of Korea}

\affil[*]{byoungho@snu.ac.kr}

\begin{abstract}
Since the history of display technologies began, people have dreamed an ultimate 3D display system. In order to get close to the dream, 3D displays should provide both of psychological and physiological cues for recognition of depth information. However, it is challenging to satisfy the essential features without sacrifice in conventional technical values including resolution, frame rate, and eye-box. Here, we present a new type of 3D displays: tomographic displays. We claim that tomographic displays may support extremely wide depth of field, quasi-continuous accommodation, omni-directional motion parallax, preserved resolution, full frame, and moderate field of view within enough eye-box. Tomographic displays consist of focus-tunable optics, 2D display panel, and fast spatially adjustable backlight. The synchronization of the focus-tunable optics and the backlight enables the 2D display panel to express the depth information. Tomographic displays have various applications including tabletop 3D displays, head-up displays, and near-eye stereoscopes. In this study, we implement a near-eye display named TomoReal, which is one of the most promising application of tomographic displays. We conclude with the detailed analysis and thorough discussion for tomographic displays, which would open a new research field. 
\end{abstract}
\begin{document}

\flushbottom
\maketitle
%
%
\thispagestyle{empty}

Since the history of display technologies began, people have dreamed to implement an ultimate three-dimensional (3D) display system. The ultimate 3D displays may provide immersive and realistic experiences, which make users hardly recognize that their experience is not true. In order to get close to the dream for the ultimate 3D displays, we need to understand what characteristics of the real object make them look real. We should find what features should be appended for state-of-the art 3D displays to alleviate the artificiality that comes from virtual objects on the displays. The most challenging issue is how to realize the essential features without any sacrifice in the conventional technical values (e.g. resolution).

Human visual system understands the real-world and perceives depth information of 3D objects via psychological and physiological cues. The psychological cues are related to the visual effect that is usually observed in daily life, including shading, perspective, illumination, and occlusion. With the virtue of advancement in the 3D rendering and computer graphics, these psychological cues could be reproduced via ordinary 2D display panels. On the other hand, the physiological cues are referred to as physical states of two eyes and objects. in terms of convergence, accommodation, and motion parallax. In order to lead the physiological cues, we need to implement a specific display system such as light field displays \cite{hong2012integral,lanman2010content,Wetzstein2012tensor,maimone2013focus,moon2016computational,lee2016additive,jang2017retinal}, stereoscopes with focus cues \cite{akeley2004stereo,love2009high,liu2010systematic,narain2015optimal,lee2016compact,lanman2017FSD,moon2017layered,lee2017foveated}, and holographic displays \cite{yeom20153d,wakunami2016projection,li2016holographic,park2017recent,Maimone2017Holo}. 

Several display technologies have been introduced and studied to reconstruct the physiological cues. However, it has been a challenging issue to provide all of the convergence, accommodation, and motion parallax without sacrifice of the display performance. Mostly, the ability of display system to reproduce physiological cues involves sacrifice in the resolution \cite{akeley2004stereo,hong2012integral,maimone2013focus,hu2014high,narain2015optimal,moon2016computational,moon2017layered,lee2017foveated}, frame rate \cite{love2009high,liu2010systematic,lee2016compact,jang2017retinal}, viewing window \cite{yeom20153d,li2016holographic}, or eye-box \cite{wakunami2016projection,lanman2017FSD,Maimone2017Holo}. Light field displays and stereoscopes with focus cues suffer from the trade-off among the spatial resolution, depth of field \cite{Konrad2017AID}, and frame rate. Although holographic displays are relatively free from the depth of field issue, they have other limitations especially in the trade-off between the field of view and the eye-box. 

Here, we present a new type of 3D displays: tomographic displays. Tomographic displays consist of focus-tunable optics, display panel, and fast spatially adjustable backlight (FSAB). The display panel expresses 2D projected images while FSAB reconstructs depth information of the 2D images. The focus-tunable optics enables it to combine the 2D images and the depth information so that users could perceive volumetric 3D images. Tomographic displays may support extremely wide depth of field, quasi-continuous accommodation, omni-directional motion parallax, the original resolution of display panels, and full frame. Tomographic displays could have various applications including tabletop 3D displays, head-up displays, and near-eye stereoscopes. 

When coming to near-eye stereoscopes, tomographic displays could provide moderate field of view within the enough eye-box. In this study, we implement a benchtop prototype “TomoReal”, which is designed as a near-eye display. TomoReal employs a focus-tunable lens and a digital micromirror device (DMD) for focus-tunable optics and FSAB, respectively. TomoReal supports more than 80 tomographic layers within 5.5D and 0.0D, the spatial resolution of 450$\times$450, the frame rate of the 60Hz, the eye-box of 7.5mm, and the field of view of 30$^\circ$. Accordingly, tomographic displays have the overall advantages of light field displays, stereoscopes with focus cues, and holographic displays.

Along with the implementation of TomoReal, we present detailed analysis and valuable discussion for tomographic displays. First, we may optimize the rendering method for tomographic layers in consideration of the accurate focus cues. Second, tomographic displays could alleviate the specific optical aberration (i.e. curvature of field) with the modification of the rendering. Third, high dynamic range (HDR) display \cite{gotoda2010multilayer,wetzstein2011layered} is also feasible application of tomographic displays. We will conclude with the discussion about some limitations of tomographic displays such as occlusion, form factor, or cost, which would open a new research field. 


\section*{Principle}
\begin{figure}[t]
\centering
\includegraphics[width=\textwidth]{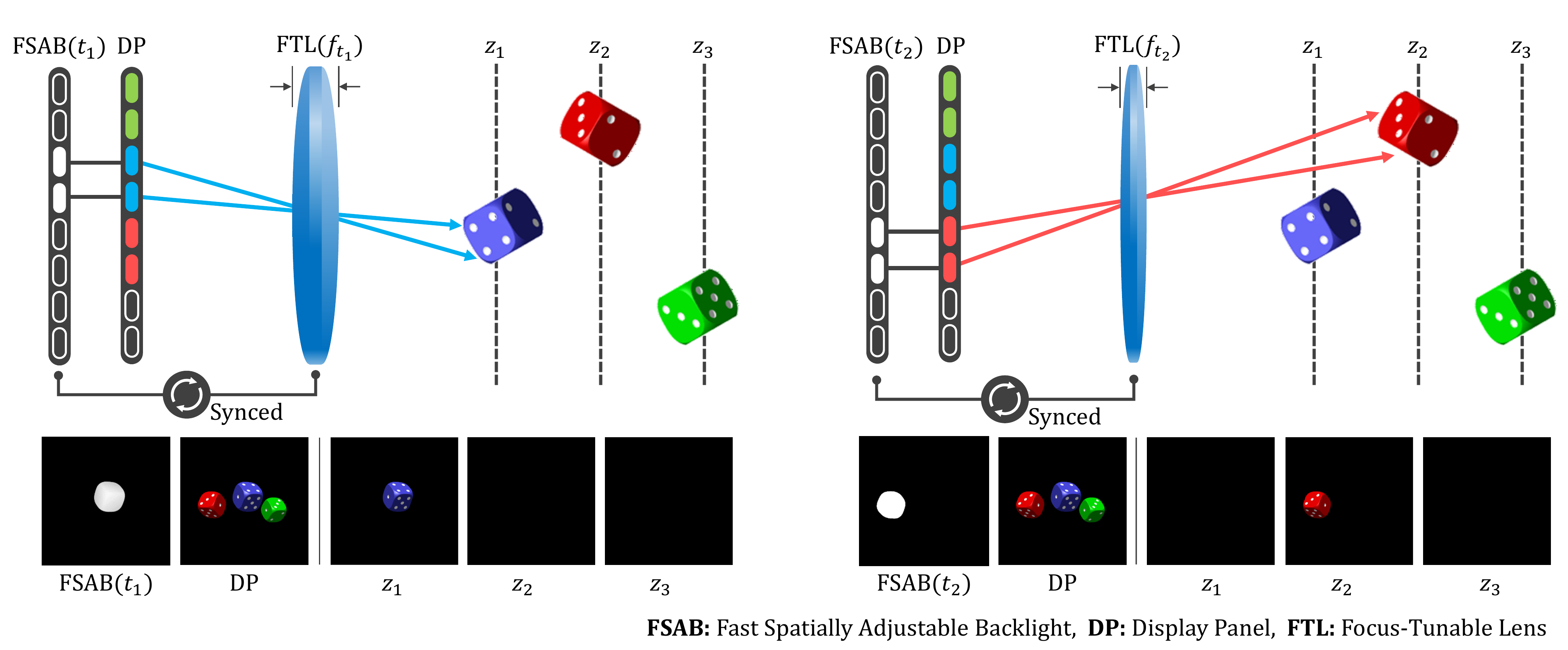}
\caption{Schematic diagram to describe the principle of tomographic displays that employ a focus-tunable lens. As shown in the figure, FSAB and focus-tunable lens are synchronized to provide users with the depth information. When the image of the display panel is formed at the depth of $z_1$, the FSAB illuminates selectively 'blue' dice of the display panel. When the focus-tunable lens forms the display panel image at the depth of $z_2$, the FSAB illuminates selectively 'red' dice of the display panel. As a result, 'blue and 'red' seem to be floated at the depth of $z_1$ and $z_2$, respectively. Note that each depth can be negative value, which means the image is created at the left-side of the focus-tunable depth.}
\label{fig:principle}
\end{figure}

The core idea of tomographic displays is in the synchronization of focus-tunable optics (e.g. focus-tunable lens or motorized stage with a lens) and fast spatially adjustable backlight (FSAB). The combination of the two elements allows an ordinary display panel to express depth information without the loss of resolution or frame rate. The process how this display system supports depth expression is as follows. FSAB illuminates a display pixel at the particular moment when the focus-tunable optics form an image of the pixel at the desired depth. In order to image all individual pixels at different depths, we employ the temporal multiplexing method. It could be considered as the transformation of the planar display panel into the arbitrary shape with the curvature.

Figure ~\ref{fig:principle} illustrates the procedure for reconstruction of 3D volumetric objects via tomographic displays. Note that the focus-tunable optics is represented by a focus-tunable lens for intuitive illustration. In the single cycle, the focus-tunable lens modulates the focal length so that the images of pixels scan along the specific range of the depth. At the same time, FSAB determines the depth information of each pixel by illuminating at the appropriate moment. FSAB projects a binary image sequence onto the display panel, which is synchronized with the focus-tunable lens. The binary image sequence is derived from the depth information of the 3D volumetric objects. In summary, while the display panel performs the common role to reproduce a 2D image that includes color and brightness, focus-tunable lens and FSAB let the 2D image have the depth information.

Synchronization with a display module and focus-tunable optics is not a new idea \cite{love2009high,liu2010systematic,hu2014high,narain2015optimal,lee2016compact,Konrad2016,lanman2017FSD,Konrad2017AID}. However, previous methodologies have been suffered from a significant trade-off among spatial resolution, accuracy of focus cues, and frame rate. In the previous approaches, display panels are updated to express each depth layer. For instance, a display panel (144-180Hz) may optimally reconstruct 3 depth layers by sacrificing the frame rate. Only with 3 depth layers, however, it is hard to cover wide depth of field while providing continuous focus cues. There could also be the resolution loss if each layer is separated by 0.6D to provide optimal range of focus cues \cite{narain2015optimal}. Therefore, it is important to break the trade-off problem for commercialization. Tomographic displays could be a competitive solution for this challenging issue.

\section*{Results}

\subsection*{TomoReal: Tomographic Near-Eye Displays for Virtual Reality}
\begin{figure}[t]
\centering
\includegraphics[width=5.0in]{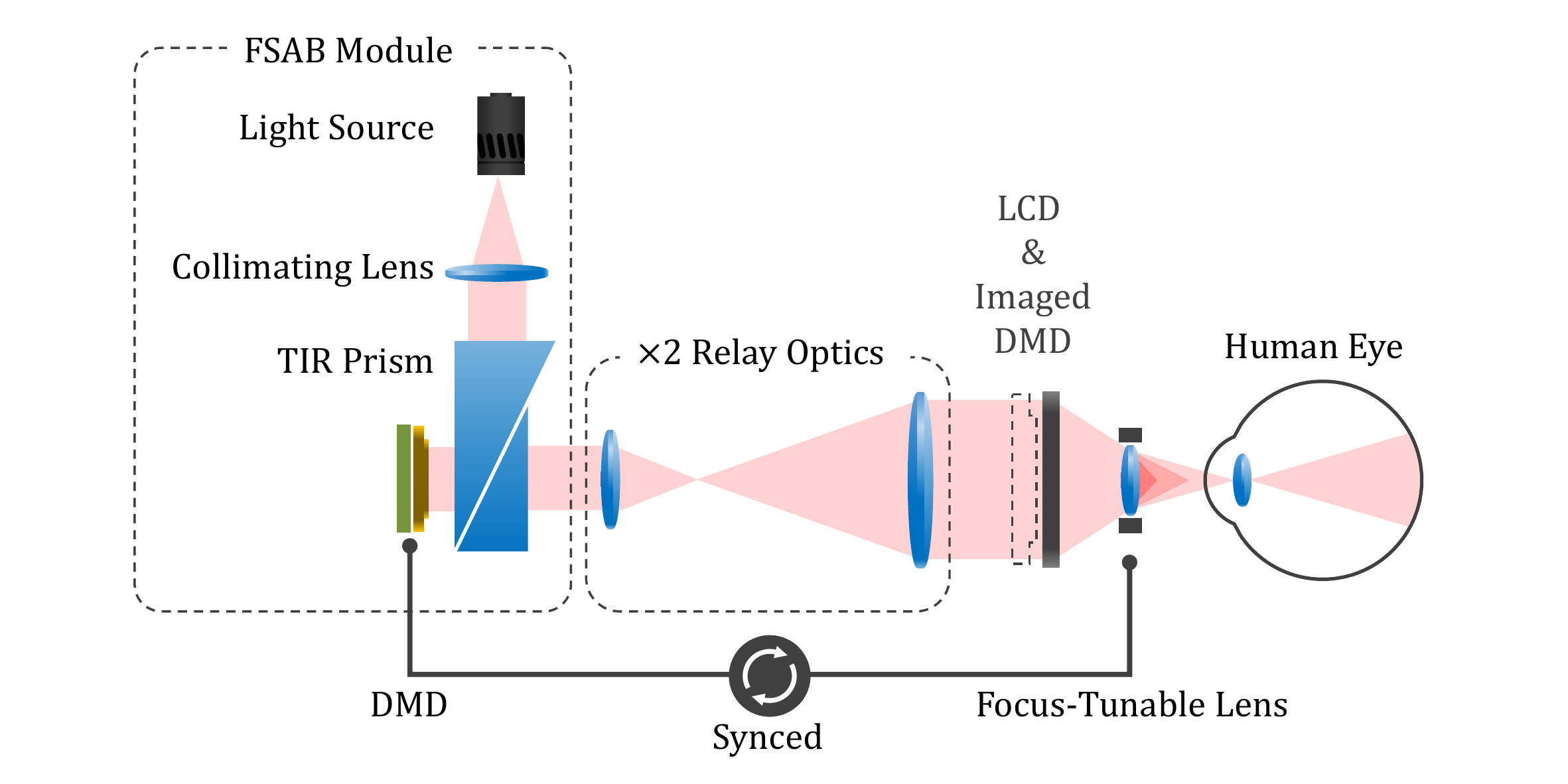}
\caption{Implementation of TomoReal using a DMD projection system and a focus-tunable lens. Relay optics may create a magnified real image of the DMD screen at the LCD plane. As the DMD projection could be updated at a fast frame rate (80$\times$60Hz), the real image of the DMD is referred to as the FSAB. The DMD is synchronized with the tunable lens (60Hz) to provide depth information. The DMD projection system could be replaced by a LED array backlight \cite{huang2017mixed}, which has advantages in the form factor. Note that the eyepiece is not illustrated in this figure. Detailed specifications of TomoReal are presented in Supplementary Material.}
\label{fig:TomoReal}
\end{figure}

We implement a prototype for tomographic displays, which we call TomoReal. TomoReal is designed as near-eye displays for virtual reality. Figure ~\ref{fig:TomoReal} illustrates schematic diagram of TomoReal. In details, TomoReal is divided into four parts: DMD projection system for the FSAB, liquid crystal display (LCD) module, focus-tunable lens, and eyepiece. The DMD projection system consists of a LED light source with a collimating lens, a total internal reflection (TIR) prism, a DMD, and relay optics with the magnification. The binary image at the DMD is magnified and projected onto the LCD module, which corresponds to the FSAB. Note that DMD could update binary images more than 100 times during 1/60 second. The eyepiece secures enough eye relief \cite{Konrad2017AID} for observer while maintaining the optimal field of view that can be achieved with the focus-tunable lens.

\begin{figure}[htp]
\centering
\includegraphics[width=6.3in]{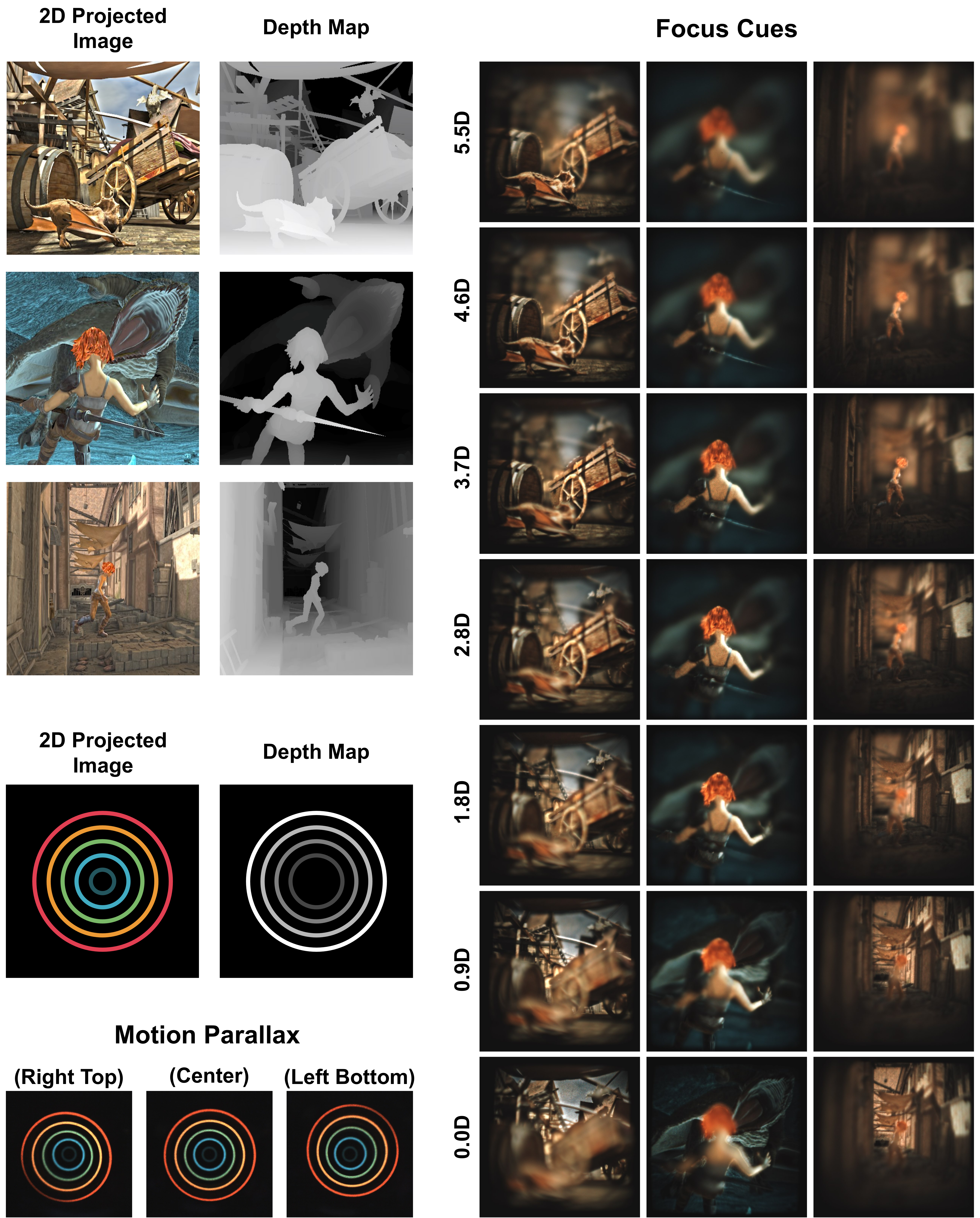}
\caption{Experimental results to demonstrate display performance of TomoReal. On the left hand side, 2D projected images and corresponding depth maps are illustrated. The source of 3D contents are from the work of Burtler et al. \cite{Butler:ECCV:2012}. On the right hand side, experimental results are demonstrated. We capture 7 photographs with different focal depths of a CCD camera. As we can see in the results, TomoReal may support quasi-continuous focus cues while preserving high resolution and contrast. On the left bottom side, we demonstrate a brief experiment to show motion parallax provided by TomoReal. Additional results and video are available in Supplementary Material.}
\label{fig:experiment}
\end{figure}

Figure ~\ref{fig:experiment} demonstrates display results of TomoReal. We employ three different 3D contents \cite{Butler:ECCV:2012} that may show significant variation in the depth. 2D projected images and depth maps are used to generate tomographic layer images. As we can see in the results, the depth information of each 3D content is well reconstructed via TomoReal. Note that TomoReal supports the original resolution of display panel with full color expression. 80 tomographic layer images are floated between 5.5D and 0.0D so that each layer is separated by 0.06875D. Note that this separation is narrow enough to provide users with quasi-continuous focus cues \cite{campbell1957depth}. We may observe clear focus cues and blur effect of reconstructed 3D contents. Note that motion parallax within the eye-box is also achieved via TomoReal as shown in the results. TomoReal provides the diagonal field of view of 30$^\circ$ within the eye-box of 7.5mm. These specifications are verified by using optical simulation tool Zemax, which could be consulted in Supplementary Material.

\subsection*{Optimal Tomographic Mapping}
In tomographic displays, each pixel of the display module would be illuminated during the specific moment. Note that the illumination time for each pixel is important since it determines brightness, contrast, and resolution of the displays. When the illumination time is too short, tomographic displays may suffer from the low brightness and contrast. If the illumination time is too long, the images of pixels are extended along the depth direction, which may degrade the resolution and obscures the focus cues. Therefore, we should optimize the illumination time for each pixel in order to find the agreed point that ensures the enough brightness, contrast, resolution, and accuracy of focus cues.

\subsubsection*{Fundamental Architecture for Optimization}
Our goal is to minimize the visual difference between target 3D objects and reconstructed scenes. We note that the depth fidelity of each pixel in tomographic displays is spatially invariant and independently determined by illumination strategy. Accordingly, the visual difference could be estimated by using incoherent point spread function (PSF) \cite{goodman2005introduction} without loss of generality. For instance, we suppose a point light source located at the desired depth $z_d$, and derive its PSF according to accommodation plane depths of $z_1^s, ..., z_m^s$. The set of PSF, $h(z_1^s, z_d), ..., h(z_m^s, z_d)$, is considered as the ground truth, which is desired to be reconstructed by tomographic displays. Thus, we may estimate the visual difference via comparison between PSF and intensity profiles reconstructed by tomographic displays. 

We employ the visual difference as the cost function $J$ in order to find the optimal illumination strategy via numerical approaches. The cost function $J$ given by
\begin{equation}
\centering
J = \sum\limits_{i = 1}^m {{{\left\| {H(z_i^s,{z_d}) - P(z_i^s)} \right\|}^2}},
\label{Eq:costfunc}
\end{equation}
where $H(z_i^s, z_d)$ denotes Fourier transform of PSF when the depth of object and image planes are $z_i^s$ and $z_d$, respectively. $P(z_i^s)$ is Fourier transform of the reconstructed intensity profile at the depth of $z_i$. The reconstructed intensity profile is determined by the sum of PSF corresponding to each tomographic layer as described by following equation.
\begin{equation}
\centering
P(z_i^s) = \frac{1}{A}\sum\limits_{j = 1}^n {{b_j}H(z_i^s,z_j^t)}, \qquad A = \sum\limits_{j = 1}^n {{b_j}},
\label{Eq:tomopsf}
\end{equation}
where $z_1^t, ..., z_n^t$ is the depth of tomographic layers whose pixel is illuminated by corresponding backlight pixel. The state (on/off) of the backlight pixel is determined by binary sequence of $b = \{ {b_1}, ..., {b_m}\} $ referred to as illumination strategy. $A$ is the sum of the binary sequence $b$, which determines the illumination time during single cycle. Accordingly, we may formulate a least square problem to find the optimal illuminate strategy as follows.
\begin{equation}
\centering
\mathop {{\rm{minimize}}} \limits_b \quad \sum\limits_{i = 1}^m {{{\left\| {H(z_i^s,{z_d}) - \frac{1}{A} \sum\limits_{k = 1}^n {{b_k}H(z_i^s,z_k^t)} } \right\|}^2}}.
\label{Eq:problem}
\end{equation}

\subsubsection*{Secure Display Brightness}
As PSF from each tomographic layer is incoherently merged via addition without destructive interference, the solution of the least square problem is trivial. The optimal illumination strategy is to turn on each pixel during the time as short as possible: the illumination time $A$ should be minimized to 1, which means that the maximum brightness of reconstructed scenes is degraded by $1/m$ times. In other words, there is trade-off between the number of tomographic layers and the brightness of reconstructed scenes. When supporting 80 tomographic layers, tomographic displays may suffer from the low brightness that is 1/80 times of ordinary display panels. Also, the low brightness could be a barrier to expand dynamic range of displays. 

We note that it is important to provide users with enough brightness of images for clear and comfortable experience. In order to secure the certain level of the brightness, we may suppose the lower bound for the illumination time, $A$. The lower bound could be considered as a constraint for the least square problem. The cost function is modified to
\begin{equation}
\centering
J = \sum\limits_{i = 1}^m {{{\left\| {H(z_i^s,{z_d}) - P(z_i^s)} \right\|}^2}}  + \gamma \left[ {{A_{low}} - A} \right]\,
\label{Eq:modicost}
\end{equation}
where $A_{low}$ is the lower bound for illumination time, and $\gamma$ is a regularization parameter. For instance, $A_{low}$ is set to $[0.625m]$ when the desired brightness is higher than 5/8 times of ordinary 2D displays. 

\subsubsection*{Mitigation of DC Noise}
In Eq.~\ref{Eq:tomopsf}, we suppose an ideal environment where the backlight pixel could be completely zero. In practical condition, however, there could be the leakage between FSAB and the light source, which is referred to as DC noise. In other words, each pixel of tomographic layer is illuminated by a constant brightness even if corresponding backlight pixel is turned off. As DC noise degrades overall contrast and diminish clear focus cues of reconstructed scenes\cite{Konrad2017AID}, it is important to mitigate this artifact for immersive experience. In order to consider this element in the optimization, the equation for reconstructed intensity profiles should be modified to
\begin{equation}
\centering
P(z_i^s) = \frac{1}{A}\sum\limits_{j = 1}^n {({b_j} + c)H(z_i^s,z_j^t)}, \quad A = \sum\limits_{j = 1}^n {({b_j} + c)},
\label{Eq:modicost}
\end{equation}
where $c$ is referred to as DC noise determined by the amount of the light source leakage.

\subsubsection*{Consideration in Human Visual Characteristics}
The last point to find optimal illumination strategy is consideration in human visual characteristics. First, response of human visual system for accommodation and blur recognition is varying according to the spatial frequencies. The most sensitive region for human vision is known as spatial frequencies of 4 to 8 cycle per degree (cpd) \cite{owens1980comparison}. We can consider this characteristic by appending the contrast sensitivity function $V(f)$ introduced by Mantiuk et al. \cite{Mantiuk:2011:HCV:2010324.1964935}. The modified cost function is given by 
\begin{equation}
\centering
J = \sum\limits_{i = 1}^m V(f) {{{\left\| {H(f; z_i^s,{z_d}) - P(f; z_i^s)} \right\|}^2}}  + \gamma \left[ {{A_{low}} - A} \right].
\label{Eq:finalcost}
\end{equation}
Second, we note that the eye-lens of human visual system has some aberration \cite{ravikumar2011creating} that influences on PSF. In order to derive more accurate PSF, we may apply ordinary human eye models demonstrated by Zernike polynomials \cite{thibos2002statistical}.

\subsubsection*{Optimal Illumination Strategy}
\begin{figure}[t]
\centering
\includegraphics[width=\textwidth]{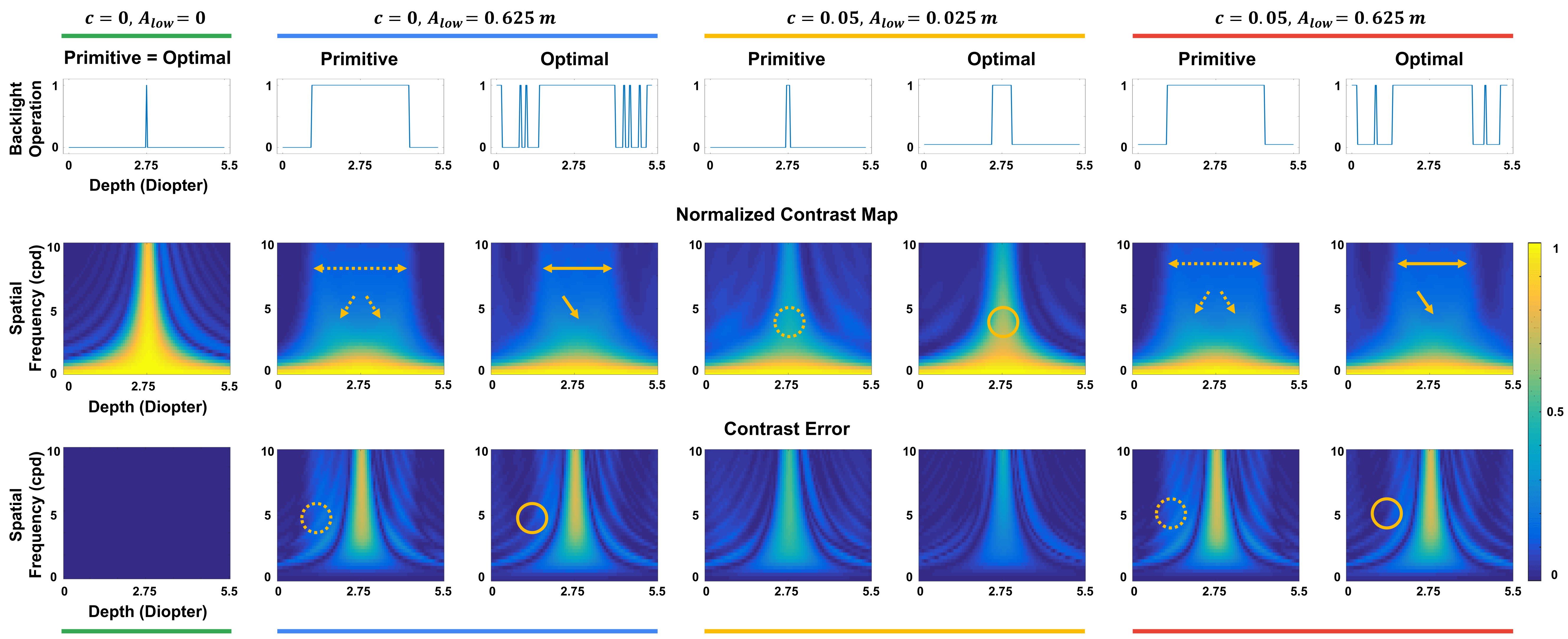}
\caption{Optimization of illumination strategy according to degree of DC noise and desired brightness: $c=0, A_{low}=0$, $c=0, A_{low}=0.625m$, $c=0.05, A_{low}=0.025m$, and $c=0.05, A_{low}=0.625m$. In the primitive strategy, each pixel is illuminated by the minimized time when its image is formed at the desired depth. In the optimal strategy, a specific backlight operation is derived by solving the least square problem. Note that the optimal strategy is identical with the primitive strategy in the naive condition ($c=0, A_{low}=0$). At the top row, we show the backlight operations according to the strategies and circumstances. At the middle row, we demonstrate normalized contrast maps that are achieved by applying corresponding backlight operation. At the bottom row, we provide contrast errors determined by the difference between the target and the reconstructed contrast maps. In some conditions, the optimal illumination strategy is to turn on backlight pixel also at the specific moment far from the desired depth $z_d$. Additional results are available in Supplementary Material. Note that TomoReal supposes the third circumstance: $c=0.05, A_{low}=0.025m$.}
\label{fig:optimization}
\end{figure}

In order to find the optimal binary sequence $b$ that minimizes the cost function given by Eq.~\ref{Eq:finalcost}, we employ the genetic algorithm \cite{whitley1994genetic}. Note that the $b_1, ..., b_m$ should be 0 or 1, which is another constraint for the optimization problem. In the genetic algorithm, the max generations and the population size are set to 1000. Figure ~\ref{fig:optimization} shows the optimization results according to degree of DC noise and desired brightness. In this simulation, we suppose that tomographic displays support the depth range between 5.5D and 0.0D with the 160 tomographic layers. The spatial frequencies of reconstructed scenes are bounded between 0 to 10cpd. The diameter of human pupil is assumed as 6mm. 

\subsection*{Advanced Applications of Tomographic Displays}
By the virtue of the remarkable capability to modulate the depth of imaged pixels, tomographic displays could have various advanced applications. For instance, tomographic display can correct optical aberration such as the curvature of field that is usually observed in near-eye display system \cite{matsunaga2015field}. High dynamic range (HDR) display \cite{gotoda2010multilayer} is also a feasible application as the intensity of the backlight could be spatially modulated. We can render HDR tomographic layer images via modification of the illumination time according to the degree of the brightness. In summary, tomographic displays have several advanced applications that provide more immersive experience.

\begin{figure}[htp]
\centering
\includegraphics[width=\textwidth]{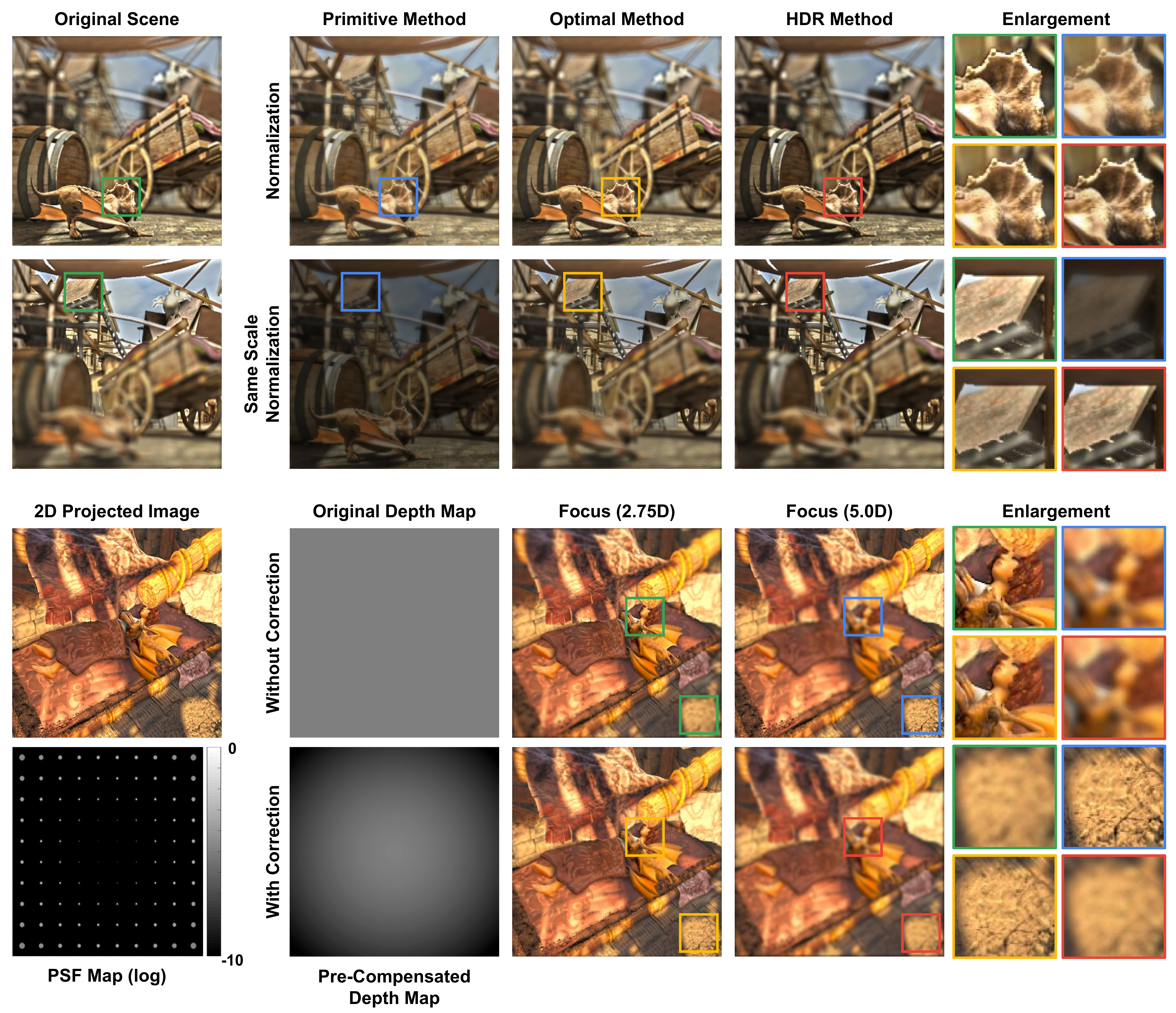}
\caption{Simulation results based on wave optics according to the illumination strategy. The number of tomographic layers is 80, and DC noise is set to 0.05. In top rows, we compare HDR method with primitive and optimal methods. HDR method shows the most convincing performance to preserve the original contrast. In bottom rows, aberration correction is demonstrated. A 3D content with a constant depth is employed to describe intuitive results. PSF map shows the optical aberration of display system: curvature of field, $10\lambda$. This aberration could be compensated by modification of the depth map. Additional results are available in Supplementary Material.}
\label{fig:advance}
\end{figure}

Figure ~\ref{fig:advance} shows simulation results that validate the proposed advanced applications, HDR displays and aberration correction, using tomographic displays. In the HDR application, the illumination time of each pixel is varying according to the desired intensity. The variation range of the illumination time lies between 0.5x and 1.5x ratio of the optimal illumination time. Second, a depth map of 3D scene is pre-compensated in order to alleviate the optical aberration (i.e. curvature of field) of display system. The pre-compensation is determined by the degree of the optical aberration. Note that we use Seidel coefficients for simulation of the curvature of field.    


\section*{Discussion}
In this study, we have explored tomographic displays and demonstrated TomoReal that is one of the most promising systems for ultimate 3D displays. There are some interesting issues and challenges for improvement in the performance of tomographic displays or TomoReal, which could be valuable discussion and future works. For instance, tomographic displays could not provide the occlusion effect in a convincing way because each tomographic layer is merged via addition. This issue could be a barrier to get closer to the dream of the ultimate 3D displays. Also, enhancement in the brightness is important if we aim to stack large amount of tomographic layers. 

Despite of the issues and challenged described above, we believe that tomographic display could be effective solution for realization of ultimate 3D displays. Most of all, tomographic displays could be modified and adopted for several applications including tabletop 3D displays \cite{lee2016additive,wakunami2016projection}, see-through near-eye displays \cite{lee2017analysis,hong2017see}, and vision assistant displays \cite{huang2014eyeglasses}. For tabletop displays, the FSAB could be implemented by using an LED array, and a liquid crystal plate with a lens array could be employed for the focus-tunable optics. For see-through near-eye displays, the eye-piece of TomoReal could be replaced by a free-form light guide to combine real-world scenes. In summary, tomographic displays will be a competitive and economic solution for all 3D technologies that aim to provide immersive experience.  

\section*{Methods}
\subsection*{Detailed Specifications of TomoReal}
For implementation of the FSAB, 1.25 inch ultra bright led spot light of Advanced Illumination is used for the illumination source. DLP9500 model from Texas Instrument is employed as the DMD, which supports full HD resolution (1920$\times$1080) and screen with 0.95 diagonal inch. The light guiding prism for DMD is customized to satisfy our specifications and convenience for experiments. We also designed relay optics that makes the real image of the DMD onto the display screen with the 2 times magnification. Note that a led array backlight is also feasible candidate for the FSAB, which has advantages in form factor compared to the DMD projection module. In Supplementary Material, we demonstrate detailed applications using the led array as FSAB.

For liquid crystal panel, we used Topfoison TF60010A model whose backlight module is eliminated. This panel may support high resolution images of 491 DPI. Focus-tunable lens (EL10-30-TC-VIS-12D) of Optotune is selected for the focus-tunable optics, which provides the wide focus tuning range between 8.3D to 20D at the real-time operation (60Hz). Using this lens, tomographic displays may have depth of field between 10.5D and 0D while TomoReal only supports 5.5D due to the camera specifications. Note that we append the eyepiece module for TomoReal in order to retain the enough eye relief for photographs. The eyepiece module consists of two identical camera lenses (Canon EF 50mm f/1.8 STM) for the 4f relay system. 

\subsection*{Synchronization of Fast Spatially Adjustable Backlight and Focus-Tunable Optics}
As demonstrated in the previous section, we used a DMD module and a focus-tunable lens for FSAB and focus-tunable optics, respectively. In order to synchronize these two modules (i.e. DMD and focus-tunable lens), Data Acquisition (DAQ) board from National Instrument is used to generate the reference clock signals. There are two different signals generated by DAQ board: one is for the tunable lens and the other is for DMD. These two different signals could be synchronized by using LabView. For the tunable lens, the triangle wave at 60Hz is generated to modulate the focal length as we desired. For the DMD, the square wave at 60$\times$80Hz is generated to update the sequential backlight images on DMD. More intuitive description of the synchronization is presented in Supplementary Material. 

\subsection*{Optimal Rendering for Tomographic Layer Images}
We can render tomographic layer images by using a 2D projected image of a volumetric object with its depth map. According to the optimal backlight operation determined by solving the least square problem, the tomographic layer images could be derived. DC noise of TomoReal is estimated as 0.05, which is optimally mitigated by the illumination time of 10/480 seconds. The illumination time corresponds to the 10 tomographic layers, which means that FSAB illuminates each display pixel by 10 times during the single cycle. Image processing tool to render layer images is implemented on Matlab.

\bibliography{main}

\end{document}